\documentclass[aps,prx,reprint,groupedaddress,superscriptaddress,nofootinbib,twocolumn,notitlepage]{revtex4-1}

\usepackage{amssymb, amsmath, amsthm, amsfonts}
\usepackage{graphicx}
\usepackage{color}
\usepackage{mathrsfs}
\usepackage{physics}
\usepackage{tikz}
\usepackage{pgfplots}

\usepackage[UKenglish]{isodate}
\usepackage[UKenglish]{babel}

\begin{document}

\title{Characterising the Non-Equilibrium Dynamics of a Neural Cell}

\author{Dalton A R Sakthivadivel}
\email{dalton.sakthivadivel@stonybrook.edu}
\affiliation{Department of Biomedical Engineering, Stony Brook University, Stony Brook, New York, 11794-5281}


\date{\today}

\begin{abstract}

We examine the dynamical evolution of the state of a neurone, with particular care to the non-equilibrium nature of the forces influencing its movement in state space. We combine non-equilibrium statistical mechanics and dynamical systems theory to characterise the nature of the neural resting state, and its relationship to firing. The stereotypical shape of the action potential arises from this model, as well as bursting dynamics, and the non-equilibrium phase transition from resting to spiking. Geometric properties of the system are discussed, such as the birth and shape of the neural limit cycle, which provide a complementary understanding of these dynamics. This provides a multiscale model of the neural cell, from molecules to spikes, and explains various phenomena in a unified manner. Some more general notions for damped oscillators, birth-death processes, and stationary non-equilibrium systems are included.

\end{abstract}

\maketitle


\section{Introduction}



The neurone, like many other natural systems, is an open system in constant exchange with its environment. Due to its small scale and the precise degree of control which a cell exerts over its internal environment, as well as the complex dynamics exhibited by neural cells in particular, it is an object of interest from a physical point of view. Neural spikes are both a regular, periodic phenomenon, and a highly complex, non-equilibrium process. As an example of self-organisation, neural firing emerges from complex but quantifiable dynamics, here involving ionic equilibria and membrane selectivity \cite{HH1952}. The neurone is surrounded by ions in its extracellular fluid, meaning it is subject to diffusion of these ions across its cell membrane, through ion channels. It maintains a negative resting potential of $-70$ mV, which requires active transport of positive $\text{Na}^+$ ions out of the cell by pumps. This results in a persistent concentration gradient, which is precisely what allows a spike to occur. When a critical voltage is reached, previously closed voltage-gated ion channels open. The positive $\text{Na}^+$ ions flow along this concentration gradient through these now open channels, leading to an upwards spike in voltage. 

Likewise, biological phenomena are often due to equally complex physical processes, such as critical or non-equilibrium dynamics, making the physics present of interest to the biologist \cite{biolCritBialek}. In neuroscience this is especially true, as while our understanding of the biology of the action potential is relatively complete, most mathematical models that explain these are built phenomenologically as non-linear oscillators that lack explanatory power \cite{neural-models}. As such, certain processes like the presence of spiking behaviour, the shape of the action potential, or the nature of bursting activity are not explained in a mathematically mechanistic way. This requires tools from complex systems theory or non-equilibrium statistical mechanics, which are uniquely meant for studying such systems \cite{WolfE8678, bialek2012}. These methods are very powerful, but often poorly understood. 


We will analyse the neurone as both a physical and biological object, explaining the fundamental phenomenology of neural dynamics with fundamental physics. Phase planes are useful objects of study as they allow us to `geometrise' the dynamics of a system. As such, we begin by considering a \emph{flow in a state space} as an analogy which will allow us to describe the system. We can model the state of the neurone as a position, a change in state as a velocity, and a change in that as a force acting on the system. In this way, we will formulate an equation of motion for the system based on $\mathbf x, \dot{\mathbf x},$ and $\ddot{\mathbf x},$ which contain $x$ and $y$ dynamics for a spiking and adaptation variable, respectively. These positions in state space depend on quantities associated to real space, which we denote as $\mathbf{r}$ such that we have $\mathbf x(\mathbf{r},t)$. We can now evaluate the dynamical and statistical behaviour of the system by analysing its motion in state space. 

\section{Main Results}

\subsection{Non-Equilibrium Resting State Dynamics}\label{TwoA}

We begin by formulating a model of resting state neural dynamics, and use this to investigate non-equilibrium phenomena like spiking in the neurone. Neural dynamics at rest are primarily characterised by a stationary state far from equilibrium. The neurone rests at an unvarying voltage of approximately $-70$ mV, despite the equilibrium concentration of the ions surrounding it being higher. This will help draw a picture of the dynamics at hand. 

As we began with, we will use some basic analogies for these quantities; primarily, that the position in state space should be a membrane voltage, because that is how a neural spike is typically measured. If the position $\mathbf x$ is the voltage inside the neurone, it is characterised by the concentration of particles at a point $\mathbf{r}$ compared to an equilibrium value, so that
\begin{equation}\label{volt-part-dens}
\mathbf{x}(\mathbf{r},t) = p(\mathbf{r},t) - p_{\text{eq}}
\end{equation}
where $p(\mathbf{r},t)$ is the density of particles, defined by the concentration of $\text{Na}^+$ at each point $\mathbf{r}$. As such, globally, $p(\mathbf{r},t) - p_{\text{eq}}$ is a gradient of ion concentrations in real space, 
\begin{equation}\label{pos-conc-grad}
\mathbf{x}(t) = \nabla [\text{Na}^+]_t
\end{equation}
which can vary in time. Clearly, the voltage is proportional to the out-of-equilibriumness of the system, implying that voltage is zero when the gradient is zero and the system is at equilibrium; this also follows from (\ref{volt-part-dens}).

To derive the force in the phase plane, $\ddot{\mathbf x}(\mathbf{r},t)$, and thus an equation of motion, we consider a slightly simplified model of neural ion dynamics. The neurone experiences diffusion of ions across its membrane due to an imbalance of charge. Therefore, this diffusion is exactly proportional to how `non-zero' $\mathbf{x}$ is, which we measure by distance in the phase plane, $\| \mathbf{x} \|$. The voltage can be thought of as a potential driving the diffusion into the cell. 

Diffusion, then, is the spatial change in local particle densities, or movement in the phase plane; in turn, it is a force acting on the phase plane position, against the gradient that defines position. This also suggests that it is the second order \emph{temporal} change in concentration gradient, which is confirmed by noting that the speed of the movement of ions is attenuated by diffusion. Together, these facts give us an important dynamical equation, that
\begin{equation}\label{force-norm-diffusion}
\ddot{\mathbf{x}}(\mathbf{r},t) = -\vec{a}\,\| \mathbf{x}(\mathbf{r},t) \|
\end{equation}
for some parameter $a$ which may vary for $x$ and $y$ in $\mathbf{x}$.

Equivalently, we have the relationship 
\[
\ddot{\mathbf{x}}(\mathbf{r},t) = -\pdv{V(\mathbf{x}(\mathbf{r},t))}{\mathbf{x}}, 
\]
where this relationship, along with (\ref{pos-conc-grad}) and (\ref{force-norm-diffusion}) imply that the function for the concentration of $\text{Na}^+$ ions is a potential for the system in state space. This is confirmed by a global continuity condition, where the number of particles travelling into the neurone (a kinetic energy) and the number of particles inside the neurone (the potential energy) are, together, conserved. Thus, we have the beginnings of a dynamical analogy by way of our definition of position. 

We note above that the movement of ions is the first order temporal change in concentration gradient $\mathbf{x}(\mathbf{r}, t)$, which gives us $\dot{\mathbf{x}}(\mathbf{r}, t)$, the movement of ions at place $\mathbf{r}$ and time $t$ with speed $\ddot{\mathbf{x}}(\mathbf{r}, t) \propto {\mathbf{x}}(\mathbf{r}, t)$. Now, since the ionic concentration is a potential energy, transferring potential energy to kinetic energy as we move towards equilibrium increases velocity. In diffusion, the final velocity is constantly zero at equilibrium, making it a point of minimum total energy. Note that the final velocity is zero in this case, unlike in an oscillator where it is constant but non-zero. This difference will be crucial to constructing the difference between a non-oscillatory stationary state and a spike, as it suggest the diffusion at equilibrium is a damped oscillator. This damping cannot be derived from the potential, and must be added heuristically, like so:
\begin{equation}\label{damp-osc-diff-eq}
\ddot{\mathbf{x}}(\mathbf{r},t) = -b \dot{\mathbf{x}}(\mathbf{r}, t) -\vec{a}\,\| \mathbf{x}(\mathbf{r},t) \|
\end{equation}

This idea is made more particular by the existence of a negative concentration gradient, which is given by the fact that there are more positive ions outside the cell than inside. The neural resting potential, or system equilibrium $\mu$, is less than the more `natural' resting position of the system, the true equilibrium $\mathbf{x^*}$. Since $\mu < \mathbf{x^*}$, ions diffuse into the cell, as diffusion adds energy to the system to correct the charge imbalance and attain equilibrium. Thus, the force acts against the direction of the gradient, and contains a small negative parameter with respect to the position. This parameter could represent spatial diffusion rate, or how many molecules pass into the membrane for a given gradient size. As stated, we will assume that $a$ is temporally constant, but allow it to vary spatially, e.g., with respect to $x$ or $y$. 

Physically, this is a cold system coupled to a high temperature bath, where a tendency to equilibrate to $\mathbf{x^*}$ drives the system. These are energy fluctuations, due to noisy particle diffusion at short spatial and temporal scales, and a potential energy from the existence of a gradient at longer scales. The difference in local and global dynamics is reflected in equation (\ref{volt-part-dens}) as compared to (\ref{pos-conc-grad}). This suggest a wealth of other information about this system, such as the possibility of defining a Fokker-Planck equation for (\ref{volt-part-dens}), and using a principled coarse-graining like the Mori-Zwanzig projection to go from this to Langevin dynamics. We will return to some of this later.

As stated, this diffusive tendency would, ordinarily, also be responsible for stopping the current, as the concentration of ions will eventually equilibrate. Eventually the gradient in (\ref{damp-osc-diff-eq}) should go to zero and thus every scale of the dynamics themselves. Additionally, once at equilibrium, these fluctuations would be countered by an equivalent dissipation that maintains equilibrium. Again, this is reflected in the fact that the force in (\ref{damp-osc-diff-eq}) is proportional to the current $\dot{\mathbf{x}}$, such that $\ddot{\mathbf{x}}$ defines a damped system.

However\textemdash as we know, the system equilibrium is maintained at $\mu < \mathbf{x^*}$. This stationary non-equilibrium point means there is a constant concentration gradient, and so (\ref{damp-osc-diff-eq}) is never zero; and yet, despite this, $\mu$ is a stationary point. Clearly there is another component in the force on the system. Recalling the membrane dynamics introduced earlier, we know that what maintains this concentration gradient is the constant action of pumps, which remove diffusing particles. As such, $\|\mathbf{x}\|$ is identically non-zero. This concentration gradient is what allows for firing to occur. 

The fact that this diffusion is, by construction, constant in time, and the dynamics stationary, means that there is damping as given in (\ref{damp-osc-diff-eq}); however, it is not associated with diffusion. The loss of kinetic energy as we move to equilibrium does not occur, because we are out of equilibrium. Instead, we have an artificial damping, due to the activity of pumps. This makes the force due to the concentration gradient decouple from the movement of ions, as both are modulated more directly by pump activity. Now, the force from displacement is independent of the effective ion movement.

The effective ion movement, in turn, is a competition between ions entering through channels and being pumped out. We can define the concentration of ions in the cell as the following equation:
\begin{equation}\label{lambda-zero-eq}
\ddot{\mathbf{x}}(\mathbf{r},t) = -\lambda_0\dot{\mathbf{x}}(\mathbf{r},t)+c(t),
\end{equation}
which defines the relationship between ions being pumped out of the cell and ions entering the cell. The pump activity is negative due to the loss of energy to the concentration gradient, while the force due to ion entrance is adding energy and thus is positive. the concentration gradient is negative because the force it exerts also requires energy loss when above equilibrium, and in a self correcting way, when it is below equilibrium it adds energy.
 
We absorb the two into an effective ion movement, which requires the pump activity to dominate movement. In other words, supposing that $\dot{\mathbf{x}} \propto c(t),$ and $| \lambda_0\dot{\mathbf{x}}(\mathbf{r},t) | > c(t)$, we can define damping as an effective movement due to pumps. Regarding any entrance of ions as a momentary perturbation,
\[
\ddot{\mathbf{x}}(\mathbf{r},t) = -\lambda\dot{\mathbf{x}}(\mathbf{r},t),
\]
from which we have the following:
\begin{equation}\label{three-comp-damp-osc}
\ddot{\mathbf{x}}(\mathbf{r},t) = -\lambda\dot{\mathbf{x}}(\mathbf{r},t) -\vec{a}\,\| \mathbf{x}(\mathbf{r},t) \|.
\end{equation}

Equation (\ref{three-comp-damp-osc}) comes from combining our effective movement with the force from the concentration gradient, as laid out in the previous justifications. In the stationary case, this implies $\dot x \propto x$, a classic fluctuation-dissipation relation; it also gives us a linear response dynamic.

Statistically, this system is a mean reverting system where damping is always in place, and when the system is excited by external perturbation, eventually it returns to $\mu$. This is shown in the figure below, the plotted solution to (\ref{three-comp-damp-osc}) for the one-dimensional case, focussing on the spiking variable. Here, a kind of `dissipation kernel' acts on normal mode oscillation in the system to give the stereotypical form of the activation function. We plot the forces as defined previously, a component for the concentration of ions in the neurone, and for the movement of the state of the neurone as it seeks to return to equilibrium. These components combine additively in the differential equation and multiplicatively in its solution. At a perturbation, when lots of ions have flowed in and the concentration is high, we have a spike. As these ions are pumped out, we have the falling action or repolarisation of the neurone. When the activity of pumps coincides with the natural oscillatory behaviour of the force of displacement, we observe hyperpolarisation. Biologically, this is the pumping of extra $\text{Na}^+$ ions out of the cell, exacerbating the existing concentration gradient in the negative direction.

\begin{figure}[h!]
\includegraphics[scale=0.25]{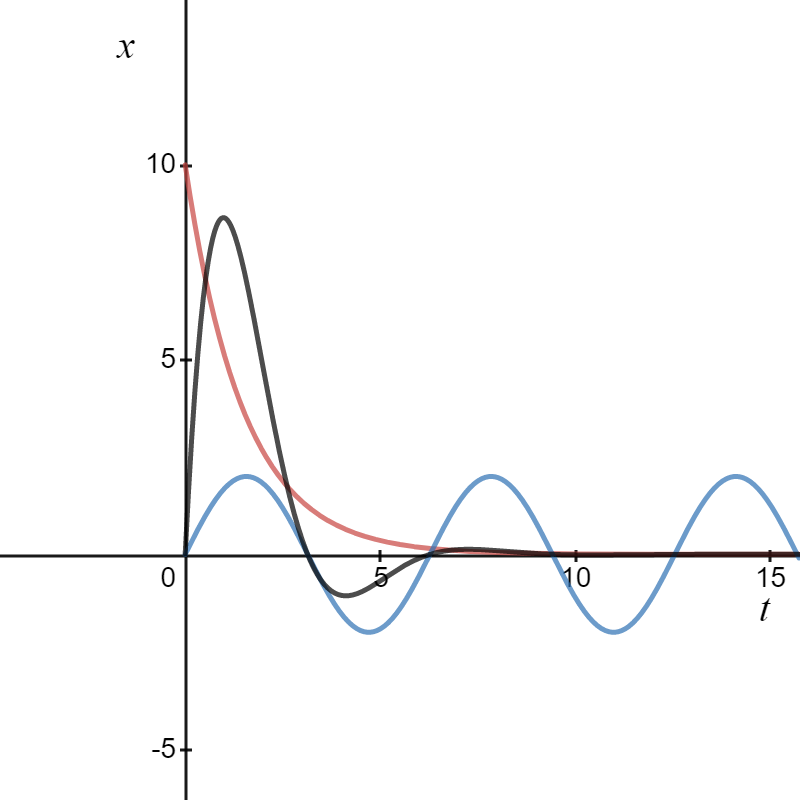}
\caption{\textbf{Stereotypical neural firing dynamics arise from the dissipation of intracellular ion concentration.} The plotted solution to (\ref{three-comp-damp-osc}) reproduces the shape of a neural action potential as a non-sinusoidal oscillation about an equilibrium. We show the solution to the single variable equivalent of (\ref{three-comp-damp-osc}) in black, the dissipative component in red, and the underlying oscillation in blue. The multiplication of this dissipation kernel with our underlying motion gives an action potential.}
\label{fig:AP-from-three-comp}
\end{figure}

We also have the suggestion of possible burst patterns in our spiking activity. Typically these are given by particular limit cycle shapes that allow for extended firing dynamics, such as a slow spiking or ion variable modulated by a fast subsystem \cite{IZH2000}. Our model is capable of exhibiting similar behaviour as the hysteresis loop in \cite{IZH2006}, where if the dissipation kernel is deactivated by setting it to a maximum in the middle of a repolarisation, the outward current is deactivated and another spike is emitted.

We will comment more on spiking in the next section.

\subsection{Transition From Diffusion to Ordered Uptake}\label{TwoB}

The dynamics at rest are only a single element of the neural system, and by far the least interesting. Neural spiking is perhaps the most famous and complex phenomenon in the human body. There is a threshold potential present in the neurone, and an all-or-nothing spiking rule that determines the transition to a spiking state. Previously, this transition has been described in a simplified way, to explain some features in artificial neural networks \cite{ANNs}. We will derive a model to describe this in a more complete way.

Spiking entails the neurone experiencing random diffusion of ions across the membrane until a critical depolarisation is reached, called the threshold potential, and then experiencing ordered, stereotypical dynamics. This could be defined as a new attractor surrounding the fixed point of the resting state. The more interesting case would be, in following with the utilisation of heat and energetic metaphors in the phase plane, modelling this phase transition from randomness to order as a Hopf bifurcation, which depends on some parameter related to the threshold potential. A limit cycle arises out of the system when this critical point is met. 

To begin, we define the Hopf bifurcation as an elimination of damping from a damped `normal form' of the system. Suppose that, following ideas from the fluctuation dissipation theorem defined previously, all asymptotic fixed points in systems capable of oscillation are actually highly damped steady-state solutions to an oscillator with underlying oscillatory dynamics. Then, the dynamics settle into the fixed point according to the following dynamical equation: 
\begin{equation}\label{gen-osc-bofx}
\ddot x = - \lambda \dot x -b(x).
\end{equation}

The characteristic polynomial of (\ref{gen-osc-bofx}) has roots $r = \frac{-\lambda \pm \sqrt{\lambda^2 - 4k}}{2}$, which is used in the solution $x = e^{rt}$. In following with the definition of the Hopf bifurcation, we will observe a periodic solution arise when the damping coefficient $\lambda$, also the real part of the roots, becomes zero. When this happens, it will render the solution
\begin{align*} 
x(t) &= e^{\text{Re}(r)t} \cdot e^{\text{Im}(r)t}\\
&= e^{\pm\sqrt{k}i},
\end{align*}
as the exponential term responsible for damping becomes one, and only the underlying oscillatory component is left. 

We would like to understand, then, how damping is driven towards zero in our system; that is, what does damping vary as a function of. In both phase transitions and Hopf bifurcations, we explore some parameter of the system that reflects the break in symmetry. The parameter of interest here is of course $\lambda$, the rate of diffusion across the membrane. Here we have a disordered phase with $\lambda < 0$ and an ordered, spiking phase with $\lambda \geq 0$. We have previously defined spiking as oscillation arising from a function for uptake, $c(t)$, countering the damping term in our dissipative dynamics. This can be thought of, in light of the above definition as leading to a limit cycle, as vectors emanating from the limit cycle centre and creating a displaced boundary. This also motivates us to define the onset of spiking activity as a phase transition, where the lowest energy configuration of the state changes along with a potential energy change. In both cases, what truly mediates the order parameter is the function $c(t)$, and so both the transition and bifurcation are due to $c(t)$.

There is a relationship between $c(t)$ and the neural state -- that is, many neurones are voltage gated. When the voltage reaches a critical point due to perturbations, called the threshold potential, $c(t)$, which is initially zero, counters the dissipative tendency $-\lambda \dot x$ in the force acting on the system. This is the discontinuity that characterises the phase transition and the critical point at which it occurs. In line with (\ref{three-comp-damp-osc}) and Figure \ref{fig:AP-from-three-comp}, when dissipation is less than or equal to $c(t),$ the concentration of ions inside the cell reaches a maximum, and a spike is created. Eventually, $c(t)$ decreases again, and dissipation dominates the ion dynamics once more. This is why despite permitting positive real root components, the oscillations are not unstable. We will make this phase transition more particular in the following section, where we use a toy model to examine exactly what causes it. 

\subsection{An Ising Model for Neural Spiking}\label{TwoC}

To investigate this phase transition, we use the Ising model, a model from statistical mechanics which is particularly adept at describing the transitions between phases. We assume the reader is familiar with the Ising model, and do not fully introduce it. Overviews are given in \cite{IsingTutorial} and \cite{Ising67}, with particular application to neural dynamics in \cite{ANNs}.

The Ising model's spin is a binarised variable taking on values $s\in\{-1,1\}$. Suppose we have $n$ channels and $m$ pumps, lying on a narrow two-dimensional lattice $\Lambda$ of size $n\cross m = N$. Suppose also that a opening in the membrane that contains an $\text{Na}^+$ ion has the value $+1$ and one that does not contain an ion is $-1$. We suppose we are close to criticality, as the resting membrane potential of $-70$ mV is close to the threshold potential $-55$ mV; this implies that a small but non-negligble proportion of spins will be $s=-1$. Indeed, we have only a small number of channels closed\textemdash but enough to destroy order in the model. 

We model spin dynamics such that $n>m$ and $n=n_\text{open}+n_\text{closed}$. This implies that when $n_\text{closed} > m - n_\text{open}$, we have dissipation due to the dominance of pumps, in line with (\ref{three-comp-damp-osc}) and our previously defined Hopf bifurcation. Similarly, if $n_\text{closed} \leq m - n_\text{open}$, the number of open channels is greater than pumps, and we have a spike; equivalently, most openings are $s=+1$ and our model approaches ferromagnetism, $\langle s \rangle \to +1$. The transition rate of this subset of spins is what we will focus on, and in particular, we will look at the feedback loop between cell voltage and opening of channels.

The low `temperature' of this Ising model suggest that we have energy diffusion, which mediates the voltage in the neurone (following (\ref{force-norm-diffusion}) as outlined in section \ref{TwoA}). The transition rate of a closed channel at $i\in\Lambda_{N}$ and time $t$ is related to the effective field around the channel, which itself is given by the proportion of closed and open channels. This is a local magnetic field, following the Glauber dynamics on the standard Ising model \cite{Glauber}. We adapt it for a dissipative non-equilibrium case, described this below. 

The transition rate obeys the following dissipative dynamic, using the local field in Glauber's Ising model, and a similar non-equilibrium Ising model described in \cite{NonEqIM, OscEqIM}:
\begin{equation}\label{IM-flip-rate}
r(s_i,t) = e^{-\lambda_0 m + \sum_j^z s_{j,t}}
\end{equation}
where $\lambda_0 m$ is the amount of outflow due to pump activity, a constant for pump rate times the number of pumps, and $z$ represents the nearest neighbours of the closed spin at $i$. The relationship suggested by this and $\lambda_0$ in (\ref{lambda-zero-eq}) indeed holds; the transition rate is sensibly proportional to the amount of damping in the system.

An incoming perturbation is a signal carrying some voltage in an adjacent neurone. This increases the local magnetic field, thus `coaxing' some channels opened according to the transition dynamics in (\ref{IM-flip-rate}). Heuristically, if this perturbation is large enough (in particular, large enough to temporarily overcome the effect of damping in the transition probability), the local field becomes so strong as to exceed the dissipation in (\ref{IM-flip-rate}) and create successive spin flips, which magnetises our Ising model. Due to the small but non-negligible positive feedback loop between the proportion of open channels and the local magnetic field on any given site $i$, the spin dynamics are critically unstable even at rest, and this provides firing in response to critical perturbations.

The oscillatory component of the action potential then comes from the fact that after opening, the transition rate for the $n_\text{closed}$ closed spins to close again remains high. They do so, and then the dissipation from pumps occurs so as to damp the system, as given in (\ref{three-comp-damp-osc}). In line with commentary at the end of section \ref{TwoA}, if this transition occurs, and then another flip back to open occurs with small probability in the middle of repolarisation, another spike will be emitted; in which case, our model will portray bursting activity.  

Finally, the probabilistic nature of (\ref{IM-flip-rate}) and the previously mentioned positive feedback loop suggest it is possible to observe spontaneous firing, although with small likelihood. This is readily justified with experimental evidence \cite{SpontFiring2004, Raman9004} and theoretical justification \cite{ChaosSpontFiring, PLOS-one-spont-firing}.

\subsection{Non-Equilibrium Spiking Steady State}

The phase transition dynamics in previous sections happen on extremely short time scales, and in reality, no neurone is quiet for long \cite{firing-rates}. We now seek a principled model of a spiking steady state, which contains the microscopic insights previously defined over a larger spatial and temporal scale. Using the non-equilibrium principles defined previously, we may describe a spiking steady state in exchange with its environment as periodic motion under the influence of various forces. By this periodicity, a limit cycle in a two-dimensional state space necessarily exists, where the spiral of states through time is collapsed into a circular shape of particular dimensions. Because the action potential has a canonical shape, this region of the phase space is well defined. We will first examine a simplified sinusoidal motion, and then address a more realistic shape corresponding to a typical action potential. 

As defined in section \ref{TwoB}, we have a limit cycle arise for specific parameterisations of our system. Over longer spatial and temporal scales, we can define this limit cycle as being due to a transformation in the dynamical quantities we measure. Much like how at macroscopic scales the statistical mechanical quantities we work with become dynamical, defined by classical theories, we make use of a change in perspective that comes with our change in scale. A direct example can be observed along the boundary of the limit cycle, where the net force equals zero, and thus the potential energy is at a minimum. In analysing this force at a macroscopic scale, we will only consider an effective maintenance of concentration gradient, that is, the net opposition to loss of energy in the system.

Noting, then, that our limit cycle has an attraction from an initial fixed point (the resting potential) to the boundary, where the potential energy is lower due to uptake dynamics inside the limit cycle, we can define a potential energy $V(\mathbf{x})$ due to ion uptake through channels: 
\[
V(\mathbf{x}) = - \int_C \mathbf{F_{up}} \cdot \dd{\gamma}.
\]
Here, we use the convention of positive forces being directed away from the limit cycle centre. 

Since the path taken, $\gamma$, is totally determined by the forces present, this line integral reduces to: 
\[
V(\mathbf{x}) = - \int_{\mathbf{x_0}}^{\mathbf{x_f}} \mathbf{F_{up}} \cdot \dd{\mathbf{x}},
\]
with $\mathbf{x_0}$ being contained inside the limit cycle and $\mathbf{x_f}$ lying on the boundary of the limit cycle. 

Keeping the metaphor, we may define the kinetic energy $T$ of a unit mass flowing along a trajectory in the state space as $\frac{1}{2} \mathbf{v}^2$. We may formulate a Lagrangian for this system in terms of $T$ and $V(\mathbf{x})$, 
\[
\mathcal{L} = T - V(\mathbf{x}) = \frac{1}{2} \dot{\mathbf x}^2 + \int_{\mathbf{x_0}}^{\mathbf{x_f}} \mathbf{F_{up}} \cdot \dd{\mathbf{x}}.
\]

We also note that there is the previously highlighted dissipation due to pump activity, and utilise the non-variational Euler-Lagrange equation, accounting for dissipation $Q$, 
\[
\dv{}{t} \pdv{\mathcal{L}}{\dot{q}} - \pdv{\mathcal{L}}{q} = Q,
\]
which yields
\[
\ddot{\mathbf{x}} - \mathbf{F_{up}} = Q.
\]

Since we know the form of this dissipation as it is given in the limit cycle, we may relate this to $Q$ and recover a complete equation of motion for the system: 
\begin{equation}\label{EOM-EL}
\ddot{\mathbf{x}} = \mathbf{F_{up}} - \mathbf{F_{diss}}.
\end{equation}
From the perspective of an initial point in the limit cycle, this dissipative component of the total force acting on the system has the natural effect of defining the boundary and restricting the final position. This also follows from the discussion of the high likelihood of `re-transition' in (\ref{IM-flip-rate}), at the end of section \ref{TwoC}.

We may also take the perspective of an initial point outside of the limit cycle; although less biologically plausible, we will note that the end result is the same. The dissipative forces present will drive the system towards an equilibrium point, a point of minimum potential energy, where there is no force acting on the system. This creates an attraction towards the limit cycle centre, and thus serves the role of a restoring force. The total force $\ddot{\mathbf x}$ is more complicated than only this, however\textemdash there remains a maintenance of ion gradients, and thus current, by the flow through channels, which resist this dissipation. 

The potential energy of the attractor is thus the integral of this restoring force. However, we must also account for the displacement of the equilibrium point by the modulation of this restoring force. We may simply add the forces inside the limit cycle such that the potential energy satisfies that it must be a minimum along the boundary of the limit cycle, and that the other forces present are accounted for: 
\[
- \int_C \left( \mathbf{ -F_{diss} + F_{up}}  \right) \cdot \dd{\mathbf{s}}.
\]
We again define the kinetic energy $T$ of as $\frac{1}{2} \mathbf{v}^2$, and with the potential energy $V$, we apply the standard Euler-Lagrange equation to find:
\[
\ddot{\mathbf x} + \mathbf{ F_{diss}} - \mathbf{ F_{up}} = 0
\]
or
\[
\ddot{\mathbf x} = \mathbf{ F_{up}} - \mathbf{ F_{diss}},
\]
recovering our equation of motion, (\ref{EOM-EL}).

Note a duality present, in which the final position defined by this minimum is now determined by the uptake force. In this case, it would be possible to define an $\mathbf{F_{net}}$ which absorbs the uptake force into the dissipative force, `shortening' the vector field to the boundary of the limit cycle and automatically restricting the final position of the system to a point along the limit cycle boundary. In particular, this is the point closest to the initial position, given by a least action principle. The role that uptake force plays in defining the limit cycle from the interior is consistent with that which is suggested by section \ref{TwoB}.

We will note one thing further about this system, namely, that current, as the flow of ions, increases with increasing imbalance\textemdash uptake or loss\textemdash of ions. Correspondingly, velocity, as the time integral of force, increases with time. Since we have projected the helical periodic motion through time onto the phase plane, here, movement through a physical distance becomes synonymous with movement in time. Thus, as a trajectory approaches the limit cycle boundary, the velocity increases in proportion to the increase in attraction, or decrease in potential energy, and the system is approximately conservative. This is true regardless of the magnitude of the vectors in time, however, we may be confident that as the system equilibriates, diffusion stops, so the force vectors decrease in magnitude as they approach the limit cycle. Nonetheless, the integral of these forces continues to increase, as they remain in the same direction in space. We notice, in line with the above observation, that this equation is Hamiltonian\textemdash that is, satisfies autonomy. This reveals a number of interesting things about the dynamics of the system, which we have used independently throughout the paper, including phase space volume preservation and the equal response of the neurone to dissipation. Thus, we have an understanding of both the consistently periodic and non-equilibrium nature of a neurone. 

As stated, we know the form of (\ref{EOM-EL}) on the microscopic, short-time scale. We can transform this to a suitable coarser scale as follows. Using (\ref{three-comp-damp-osc}) and the discussion in \ref{TwoC}, the spiking steady state condition implies that $\lambda > 0$, giving
\[
\ddot{\mathbf{x}} = \vec{\theta}\, \dot{\mathbf{x}} -\vec{a}\,\| \mathbf{x} \|
\]
for some \emph{new} effective parameter, $\vec{\theta}$. We also use the homogeneity of $\mathbf{r}$ at large spatial scales to look only at the phase plane dynamics, and therefore, decouple the phase plane variables from the probability distribution across $\mathbf{r}$.

We separate this system into a spiking variable $x$ and an adaptation variable $y$,
\begin{equation*}
\begin{aligned}
\ddot{x} &= a\dot{x} -b\| \mathbf{x} \| \\
\ddot{y} &= c\dot{y} - d\| \mathbf{x} \|
\end{aligned}
\end{equation*}
where we have rewritten $\theta_x = a$, $\theta_y = c$, $a_x = b$, and $a_y = d$. Now, we integrate with respect to time: 
\begin{equation}\label{PID-like-ODE}
\begin{aligned}
\dot{x} &= ax -b \int^s \| \mathbf{x} \| \dd{t}\\
\dot{y} &= cy - d\int^s \| \mathbf{x} \| \dd{t}
\end{aligned}
\end{equation}
with $s$ a time variable. This yields a system with the form of a PID controller, which controls motion by both position and the integral of position, confirming our notion of the maintenance of a stationary, out-of-equilibrium point. 

We may argue that the time integral of phase plane distance is area $xy$, due to the following rewriting of the system:
\[
\dd{x} = f(x,y)\dd{t} \\
\]
which implies
\[
\int^s \| \mathbf{x} \| \dd{t} = \int^s \| \mathbf{x} \| \frac{\dd{x}}{f(x,y)}.
\]

If we assume $f(x,y)$ is approximately constant over infinitesimally small time periods $\dd{t}$, we have,
\[
\int^s \| \mathbf{x} \| \dd{t} = \frac{1}{f(x,y)} \int^s \| \mathbf{x} \| \dd{x}.
\]
Performing this integral, we obtain the integral of the boundary of a curve (the distance from zero at each time) along the curve, which is the area,
\begin{equation}\label{area-integral-pre-LV}
\int^s \| \mathbf{x} \| \dd{t} = \alpha \int^s \| \mathbf{x} \| \dd{x} = \alpha xy(s).
\end{equation}

\begin{figure}[htb!]
\includegraphics[scale=0.8]{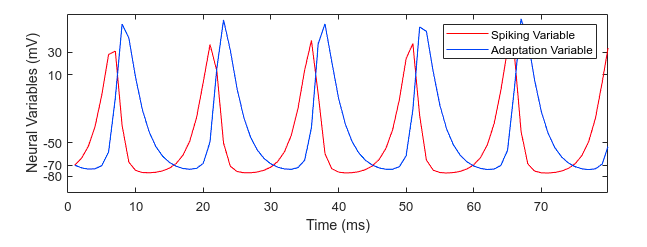}
\caption{\textbf{Stereotypical neural firing dynamics given by the numerical solution of the Lotka-Volterra equations.} Plotting a temporally discretised, numerical solution to the Lotka-Volterra solution for initial $x,y=10$, integration constant $C = -80$, $b'= d'= 0.014$, $a=0.6$, and $c=0.45$. Obvious features of the neural spike are present, such as oscillation about an equilibrium point of $\mu=-70$, shallow hyperpolarisation and large amplitude spikes, short ($<15$ ms) inter-spike interval, a consistent point at which a spike begins rising (threshold potential), and a fairly regular or stereotypical shape. These features of the equation match \emph{precisely} the features known from biological data. The physical or biological interpretation of these parameter values is not yet clear.}
\label{fig:LV-eq-matlab}
\end{figure}

Then, from (\ref{PID-like-ODE}) and (\ref{area-integral-pre-LV}), and arguing that adaptation varies in opposition to spiking due to the shared resources imposed by a local continuity condition, we have the Lotka Volterra equations
\begin{equation*}
\begin{aligned}
\dot{x} &= ax -b'xy \\
\dot{y} &= -cy + d'xy.
\end{aligned}
\end{equation*}
In fact, the solution to the Lotka-Volterra system exactly plots a stereotypical action potential for a particular parameter set, shown in Figure \ref{fig:LV-eq-matlab}. Together, these imply that the Kolmogorov process underlying the Lotka-Volterra equations are comparable to neural dynamics, which is quite reasonable, since they share the same non-equilibrium diffusion based characteristics. It has also been shown that the Lotka-Volterra equations have an associated Hamiltonian; further, since they model the interactions between predator and prey, we are inspired to refer to the dynamics of our own system as the interplay between a spike and an adaptation variable; we interpret the spike as using resources, and the adaptation variable, which lengthens the inter-spike interval, as allowing replenishment of these
resources.

\onecolumngrid

\twocolumngrid

\section{Conclusion}

We have combined dynamical and statistical insights to understand how neural spiking arises as a non-equilibrium phenomenon, and derived a model to capture these qualities. The model, from molecules to spikes, uses various dynamical quantities and their statistical mechanical underpinnings to study the evolution of neural membrane voltage. We evaluated the essential role that the non-equilibrium nature of complementary diffusion and dissipation play, at once as simple as maintaining a non-equilibrium stationary state and as complex as allowing spiking behaviour. Due to its explanatory power and insightful results, this model is a valuable description of neural cell dynamics. 

\bibliographystyle{unsrt}
\bibliography{main}

\end{document}